\documentclass[a4paper,11pt,fleqn]{iopart}
\usepackage{graphicx}
\usepackage{subfig}
\usepackage{cite}
\usepackage{color}
\usepackage{amsfonts}
\usepackage{srcltx}
\usepackage{xifthen}
\newcommand{\sig}[2]{\ifthenelse{\isempty{#1}}{\sigma_{#2}}{\sigma_{#1}^{#2}}}
\newcommand{\sigS}{\sig{s}{}}
\newcommand{\sigE}{\sig{e}{}}
\newcommand{\sigED}{\sig{e}{\dagger}}
\newcommand{\sigSD}{\sig{s}{\dagger}}

\begin{document}
\title{Oscillation and synchronization of two quantum van der Pol oscillators}
\author{Lisa Morgan and Haye Hinrichsen}
\address{Universit\"at W\"urzburg, Fakult\"at f\"ur Physik und Astronomie, Am Hubland, \\ 97074 W\"urzburg, Germany}

\ead{lmorgan@physik.uni-wuerzburg.de, \\ \hspace{13mm}hinrichsen@physik.uni-wuerzburg.de}

\begin{abstract}
The synchronization properties of two self-sustained quantum oscillators are studied in the Wigner representation. Instead of considering the quantum limit of the quantum van-der-Pol master equation we derive the quantum master equation directly from a suitable Hamiltonian. Moreover, the oscillators are coupled in incorporating an additional phase factor which shows up in the mutual correlations. 

\end{abstract}

\def\d{{\rm d}}
\def\tiny{}
\def\text{}
\def\sort{\textnormal{sort}}
\def\eigenvalues{\textnormal{eigenvalues}}
\def\({}
\def\){}
\def\fact#1{\paragraph{#1}}
\def\0{\emptyset}
\def\ket#1{|#1\rangle}
\def\bra#1{\langle#1|}
\def\braket#1#2{\langle#1|#2\rangle}
\def\ketbra#1#2{|#1\rangle\langle #2|}
\newcommand{\vect}[1]{\mbox{\boldmath${#1}$}}
\def\eqr#1{(\ref{#1})}

\def\comment#1{\color{red}[\textbf{comment: #1}]\color{black}}
\def\commentresolved#1{\color{green}[\textbf{comment: #1}]\color{black}}
\def\mark#1{\color{red}#1 \color{black}}

\pagestyle{plain}

\section{Introduction}

In physics the 'synchronization' stands for the emergence of temporally correlated behavior in composite systems. In the 20th century the main focus has been on the synchronization of periodically driven systems (see e.g. Ref.\cite{Blekhman88}). In the past two decades, however, the search for synchronization has moved to chaotic systems~\cite{Boccaletti02}. The interest in these systems stems from the fact that their temporal evolution sensitively depends on the initial condition, which is unfavorable for synchronization. Nevertheless it turns out that the trajectories of chaotic units can be synchronized under certain conditions. More recently the focus shifted again from classical to quantum systems. Here the presence of non-deterministic quantum fluctuations is expected to counteract synchronization, and therefore it would be interesting to explore under which conditions and in what sense quantum systems are able to synchronize. 

By definition synchronization is an irreversible phenomenon. Consequently, synchronization phenomena are only possible in systems with a broken time reversal symmetry. Since the standard quantum theory of closed systems is invariant under time reversal, it follows that quantum synchronization is only possible in open quantum systems that interact with the environment. On the other hand, interaction with the environment usually leads to decoherence of quantum effects~\cite{Schlosshauer2008}, and this seems to be unfavorable for synchronization. 

The most common models for the study of decoherence in open quantum systems are the harmonic oscillator and two-level spin systems~\cite{Schlosshauer2008}. Setups  where harmonic oscillators comprise both the system and the heat bath are studied under the name of \emph{quantum Brownian motion}. It was found that the form of the coupling between system and bath determines the time scale of
decoherence and the preferred states of the system (so called \emph{pointer states}). On the other hand, \emph{spin-boson models} describe situations where spin particles evolve under the influence of an harmonic oscillator bath. In this kind of models synchronization has been observed \cite{Giorgi2014}, depending on the coupling mechanisms between system and bath. The problem, however, is that the bath has a strongly dissipative influence: While two coupled oscillators will attune their frequencies over time, they will also both be damped due to the coupling to the bath.

The observation that two non-driven quantum-mechanical systems cannot be synchronized is not too surprising. In fact, even in the classical realm synchronization is only possible if we are dealing with \textit{driven} or self-propelled systems, simply because the desired synchronized state is an evolving one which depends on a continuous supply of energy to compensate dissipation. As we expect the same to be true in the quantum case, it is near at hand to look for synchronization in \textit{driven} quantum systems out of equilibrium.

The external drive could be implemented in different ways. For example, it could be introduced as a periodically changing driving force, but this would introduce an inherent time scale corresponding to the driving frequency which may interfere with other periodic properties of the system. For this reason we are interested here in freely evolving self-propelled quantum systems, which provide a mechanism to supply energy without imprinting a frequency.

One of the simplest self-sustained oscillators in classical physics is the so-called van der Pol oscillator~\cite{VanDerPol34}, which evolves according to the second-order differential equation
\begin{equation}
\ddot {\vec x} + \epsilon (1-x^2) \dot {\vec x}+\vec x = 0\,.
\end{equation}
The drive is implemented in the second term, whose influence can be interpreted as a energy-dissipating friction for $|\vec x|>1$ and some kind of energy-injecting anti-friction for $|\vec x|<1$. As a result, the dynamics of the classical van der Pol oscillator converges into a limit cycle at $|\vec x|=1$ where both influences compensate one another. When suitably coupled, a ring of such classical oscillators is known to synchronize~\cite{Nana06}. It is therefore interesting to see what happens in the quantum case.

Recently a quantum analog of the van der Pol oscillator has been studied in Refs.~\cite{Lee2013,Walter2014,Lee2014}. For example, in Ref.~\cite{Walter2014} it was found that with increasing coupling between two oscillators
the weight of the joint-probability distribution $P(x_1,x_2)$ becomes gradually concentrated near the line $x_1=x_2$, indicating phase correlations. In conclusion the authors remark that strict frequency locking (as apparent in classical systems) cannot be found in quantum systems due to quantum noise. However they state that they find strong frequency entrainment.

In this paper we extend these previous works. First we will review the master equation for the quantum van der Pol oscillator, presenting a different derivation. Having studied the time evolution of a single unit we then focus on two coupled van der Pol oscillators in the second part. In particular, we introduce a generalized coupling involving an additional phase factor and calculate the steady state solution. As usual, we visualize the synchronization properties by using Wigner functions. Moreover, we show that the visualized results are in agreement with a direct evaluation of the correlation between the oscillators. Finally we calculate the entanglement of the two oscillators.

\section{Setup of the master equation}\label{sec:General}

Let us first consider a single quantum van der Pol oscillator. Like an ordinary harmonic oscillator this system possesses a discrete energy spectrum. Because of the coupling to a heat bath, the energy levels are randomly populated. This can be described in terms of a time-dependent density matrix $\rho(t)$. For high energies, where the distribution over high energy levels is quasi-continuous, we expect the system to behave classically. Conversely, quantum effects are expected to play a role in the limit of low energies. Following Refs.~\cite{Lee2013,Lee2014,Walter2014} we will consider here the quantum limit at very low energies. In this limit the physics is dominated by the ground state and the first excited state while all higher energy levels can be neglected. This effectively leaves us with a two level system. 

The time evolution of a van der Pol oscillator in this quantum limit is given by the master equation~\cite{Lee2013}
\begin{eqnarray}
    \label{eq:MasterOneVdP}
    \dot{\rho} (t)=-i\left[H,\rho (t)\right]
    &+\kappa \left(2 \sigma ^+\rho
        (t)\sigma ^--\left\{\sigma ^-\sigma ^+,\rho
    (t)\right\}\right)\\
    &+ \lambda\left(2 \sigma ^-\rho (t) \sigma ^+-\left\{\sigma
    ^+\sigma ^-,\rho (t)\right\}\right),\nonumber
\end{eqnarray}
In this equation $\sigma^+$ and $\sigma^-$ are the creation and annihilation operators acting on the two lowest levels of the oscillator. The free system Hamiltonian is given by 
\begin{equation}
 H=\omega  \sigma ^+\sigma ^-,
\end{equation}
where $\omega$ is the oscillator frequency. Moreover, $\kappa$ and $\lambda$ are the coupling constants of the interaction with the bath.

This equation can be derived in two different ways. Following \cite{Lee2014} one can start from the general quantum master equation for van der Pol oscillators,
$\dot{\rho}=-i\left[H,\rho\right]+\kappa_1\mathcal{L}\left[a^{\dagger}\right]\rho +\kappa_2\mathcal{L}\left[a^2\right]\rho$, where $\mathcal{L}$ denotes the Lindblad superoperator.
By taking the limit $\kappa_2\rightarrow\infty$ one arrives at Eq.~(\ref{eq:MasterOneVdP}) with $\kappa=\kappa_1$ and $\lambda=2\kappa_1$.

Alternatively one can start with the Hamiltonian of a spin system in a spin bath similar to the approach in Ref.~\cite{Englert2002}. The time evolution is then obtained by assuming that the system interacts with one bath spin at a time for a short interaction period. In this approach the two terms in Eq.~(\ref{eq:MasterOneVdP}) with $\kappa$ and $\lambda$ are generated by bath spins that are initially in the ground or excited state, respectively, before the interaction takes place. In this case we find that the rate of the incoming particles in the ground and excited state can be chosen individually. This means that the ratio $\kappa/\lambda$ does not necessarily need to be $1/2$.
Moreover we see that $\kappa$ and $\lambda$ can be interpreted as the probability of finding a bath spin in the ground or excited state.
The ratio of these parameters can therefor be translated into a temperature of the bath by a Maxwell-Boltzmann factor~\cite{Englert2002}
\begin{equation}
    \frac{\lambda}{\kappa} = exp\left(-\frac{\hbar\omega}{k_B\Theta}\right).
\end{equation}
In physical terms any bath temperature is allowed in this second approach as opposed to one specific temperature in the first approach.
Technical details of the derivation can be found in~\ref{sec:SpinSpin}.

To study correlations and synchronization properties of quantum systems one needs at least two self-sustained oscillators which are mutually coupled. The corresponding master equation can be obtained by adding the contributions of both oscillators supplemented by a suitable coupling term:
\begin{eqnarray}
    \label{eq:TwoCoupledVdP}
    \dot{\rho} (t)=-i\left[H,\rho (t)\right]
    &+\kappa\sum _{i=1}^2\left(2 \sigma ^+_i\rho
        (t)\sigma ^-_i-\left\{\sigma ^-_i\sigma ^+_i,\rho
    (t)\right\}\right)\\
    &+\lambda\sum _{i=1}^2\left(2 \sigma ^-_i \rho (t) \sigma ^+_i-\left\{\sigma
    ^+_i\sigma ^-_i,\rho (t)\right\}\right)\nonumber\\
    &+V \left(2 c\rho (t)c^{\dagger }-c^{\dagger
    }c\rho (t)-\rho (t) c^{\dagger } c\right)\nonumber
\end{eqnarray}
Here the scalar $V$ determines the strength of the coupling between the two oscillators. In the following we choose $\lambda=2\kappa$, $V>0$ and $\omega$ is the same for both. For the operator $c$ occurring in the coupling term we use the choice $c=\sigma ^-_1+e^{i\phi}\sigma ^-_2$, as will be discussed in detail in section~\ref{sec:TwoCoupled}.

\section{Phase Space Representation using Wigner functions}\label{sec:Wigner}

Quantum systems described by the master equation (\ref{eq:TwoCoupledVdP}) can be realized in different physical setups. One practical and experimentally advanced realization are quantum optical experiments, where the qubits are represented by single photons. In this research area the Wigner quasi probability function is a common tool to visualize the qubit's quantum state $\hat{\rho}$ and its evolution. It also provides a convenient way to compare the quantum case with the classical behavior. Following~\cite{Leonhardt2005} we now derive a suitable form of the Wigner function.

First we define the \emph{quadratures}
\begin{equation}
\hat{q}=\frac{\sigma^++\sigma^-}{\sqrt{2}}\,, \qquad
\hat{p}=i\frac{\sigma^+-\sigma^-}{\sqrt{2}}\,.
\end{equation}
Since $\sigma^-=\frac{1}{\sqrt 2} (\hat q + i \hat p)$ it is clear that $\sigma^-$ encodes the complex amplitude of the oscillator in phase space and that the quadratures represent its real and imaginary part.  For an optical realization the quadratures correspond to the in-phase and out-of-phase component of the electric field amplitude. Since they can be regarded as  position and momentum of
an oscillator, it is therefore reasonable to represent them in a classical phase space.

\begin{figure}
  \centering
    \includegraphics[width=\textwidth]{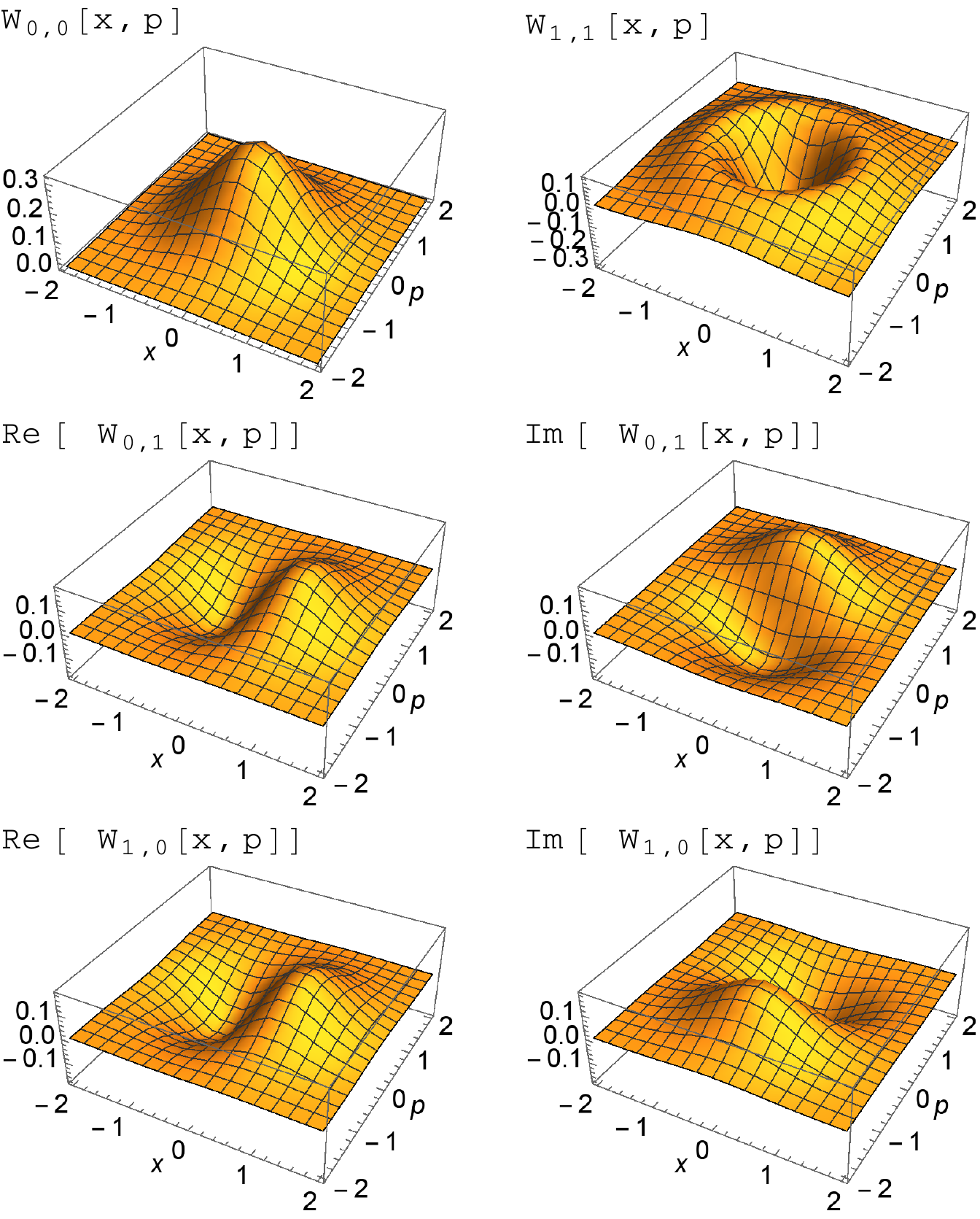}
    \rule{35em}{0.5pt}
  \caption[Wigner function components]{Components $W_{m,n}(x,p)$ of the Wigner function of a two-level system in Eq.~(\ref{FourWigner}). }
  \label{fig:WignerComponents}
\end{figure}

For a pure quantum state $|\psi\rangle$ the Wigner function is given by
\begin{equation}
    W(x,p)=\int_{-\infty}^\infty \psi^*(x+y)\psi(x-y)\,e^{2ipy}\,dx.
\end{equation}
where $\psi(x)=\langle x|\psi\rangle$ and $\hat q|x\rangle =x|x\rangle$ and consequently
\begin{equation*}
\psi^*(x+y)\psi(x-y)= \braket{x-y}{\psi}\braket{\psi}{x+y}= \bra{x-y}\rho\ket{x+y}.
\end{equation*}
Realizing that the pure state $\rho=|\psi\rangle\langle\psi|$ is mapped linearly to the Wigner function~$W$, it is clear that the same relation will hold for mixed states as well. Writing the mixed density matrix in the orthonormal energy eigenbasis $\rho=\sum_{m,n} \rho_{m,n}\ket{m}\bra{n}$, the Wigner function can be expressed as

\begin{equation}\label{eq:WignerComponents}
    W(x,p)=
    \sum_{m,n}\rho_{m,n} W_{m,n}(x, p),
\end{equation}
with
\begin{eqnarray}
    W_{m,n}(x,p)
        & = \int_{-\infty}^\infty \braket{x-y}{m}\braket{n}{x+y}\,e^{2ipy}\,\d y\\\nonumber
        & = \int_{-\infty}^\infty \phi_m(x-y)\phi^*_n(x+y)\,e^{2ipy}\,\d y.
\end{eqnarray}
where $\phi_n(x)=\braket{x}{n}$.
Using the common representation of the Heisenberg algebra ($q\rightarrow x$ and $p\rightarrow -i\hbar\frac{\partial}{\partial x}$) these wave functions are just the harmonic oscillator eigenfunctions
\begin{equation}
\phi_n(x)= \sqrt[4]{\frac{1}{\pi  4^n (n!)^2}}\; e^{-\frac{x^2}{2}} H_n(x),
\end{equation}
where $H_n(x)$ are the Hermite polynomials. This allows us to compute the first four components of the Wigner function
\begin{eqnarray}
W_{0,0}(x,p)=e^{-(x^2+p^2)} \nonumber \\
W_{0,1}(x,p)=\sqrt{2}e^{-(x^2+p^2)}(x+ip) \\
W_{1,0}(x,p)=\sqrt{2}e^{-(x^2+p^2)}(x-ip) \nonumber \\
W_{1,1}(x.p)=e^{-(x^2+p^2)} (2x^2+2p^2-1) \nonumber \,.
\label{FourWigner}
\end{eqnarray}
which are plotted in Fig.\ref{fig:WignerComponents}. As one can see, $W_{0,0}$ and $W_{1,1}$ are rotationally invariant in phase space while for the other two functions this symmetry is broken.

\section{Properties of a single quantum van der Pol oscillator}\label{sec:SingleVdP}

Let us first consider a single quantum van der Pol oscillator. Since the master equation~(\ref{eq:MasterOneVdP}) can be interpreted as a set of three linear differential equations it can be solved exactly. For $t \to \infty$ one finds that the density matrix $\rho(t)$ always approaches the steady state solution
\begin{equation}
\label{eq:SteadyState}
 \rho_{stat}=\lim_{t \to \infty}\rho(t)=  \frac{2}{3} \ket{0}\bra{0} + \frac{1}{3} \ket{1}\bra{1}.
\end{equation}
This state is a probabilistic mixture of the ground state and the first excited state. Because of (\ref{FourWigner}) the corresponding Wigner representation is symmetric under rotation in phase space.

\begin{figure}[tbp]
  \centering
    \includegraphics[width=0.9\textwidth]{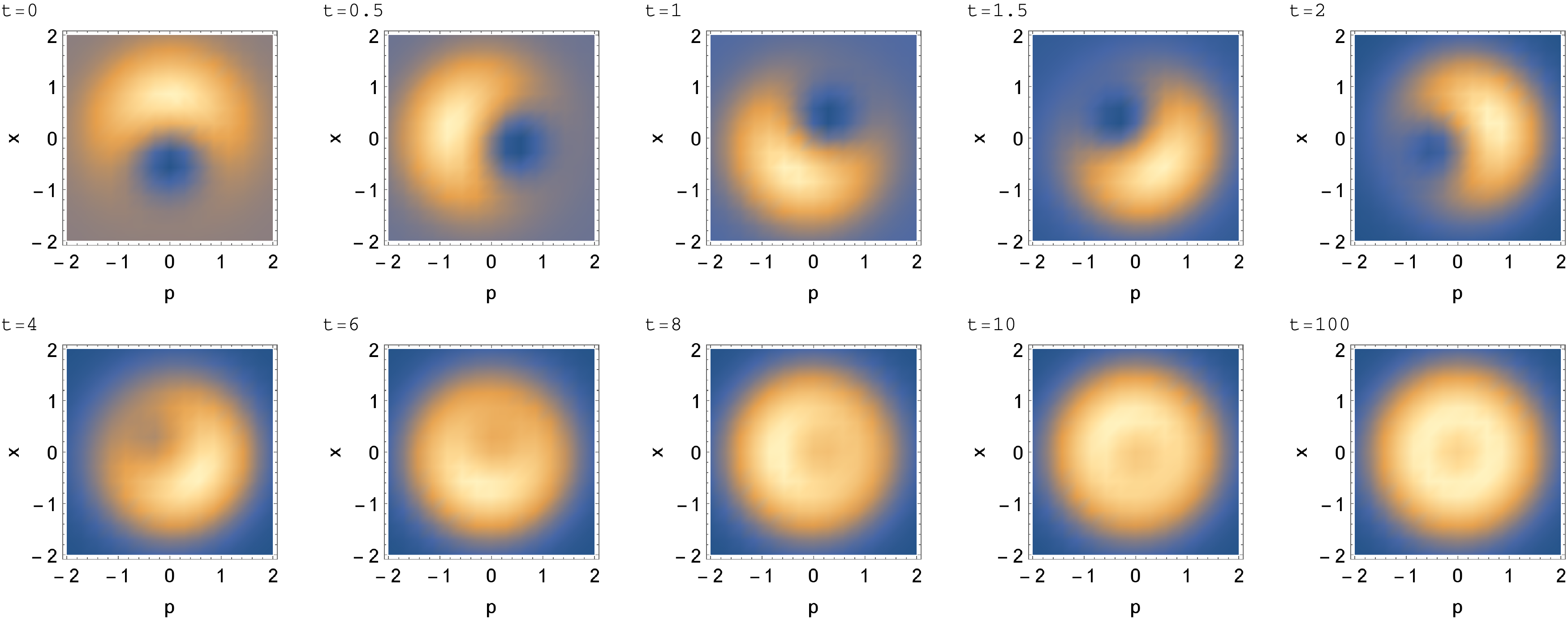}
  \caption[Time evolution of a single quantum van der Pol]
        {Wigner probability density of a single quantum van der Pol oscillator at different times.
        The initial condition is given by equation~(\ref{eq:InitialCondition}). For long times $t$
        the distribution approaches the steady state solution.}
  \label{fig:RotationDensityPlots}
\end{figure}

Turning to the time evolution it would be interesting to see the actual oscillation of the quantum oscillator in the Wigner representation. To this end we have to start with a probability distribution which is peaked at a certain point in phase space away from the origin. For example, starting with the initial (pure) state
\begin{equation}
\label{eq:InitialCondition}
\rho(0)=\frac{1}{2} \Bigl(\ket{0}-i\ket{1}\Bigr)\Bigl(\bra{0}+i\bra{1}\Bigr)\\
\end{equation}
which is not symmetric under rotation in phase space, we obtain the time evolution shown in Fig.~\ref{fig:RotationDensityPlots}. As can be seen, the maximum of the probability density rotates counter clockwise and moves to the center until the steady state distribution (\ref{eq:SteadyState}) is reached.

Why does the dynamics evolve into a rotation-symmetric probabilistic combination of $\ketbra 00$ and $\ketbra 11$? This can be explained as follows. As one can see from equation~(\ref{eq:MasterOneVdP}) the bath is coupled to the system via the creation and annihilation operators $\sigma^+$ and $\sigma^-$. These operators switch between the two levels and destroy any quantum-mechanical superposition of $\ket 0$ and $\ket 1$. Therefore the phase information is destroyed by decoherence and the system is localized in a probabilistic combination of the ground and excited state. The mixing ratio (1:2) between the two states is given by the ratio of $\kappa$ and $\lambda$ in Eq.~(\ref{eq:MasterOneVdP}).

\section{Coupled oscillators}\label{sec:TwoCoupled}

Let us now study possible synchronization effects between two coupled quantum oscillators, as described by Eq.~(\ref{eq:TwoCoupledVdP}). In the literature the operator $c$ is usually defined as $c=\sigma ^-_1-\sigma ^-_2$~\cite{Walter2014}. In the present work, however, we use the more general form
\begin{equation}
c=\sigma ^-_1+e^{i\phi}\sigma ^-_2
\end{equation}
with an additional coupling phase parameter $\phi$. The coupling operator $c$ can be interpreted as a simultaneous transition between ground and excited state of both oscillators which is induced by the heat bath. Depending on the coupling phase $\phi$ one of the oscillators acquires an additional phase factor relative to the other. As we will see below, this phase will show up in the phase probability distribution of the two oscillators in the steady state.

Again it is straight-forward to compute the steady state. It depends only on the ratio $\gamma := \frac{V}{\kappa}$ while it turns out to be independent of $\omega$:

\begin{equation}
\rho = \left(
\begin{array}{cccc}
 \frac{\gamma +3}{\gamma  (8 \gamma +27)+27} & 0 & 0 & 0 \\
 0 & \frac{(\gamma +2) (\gamma +3)}{\gamma  (8 \gamma +27)+27} &
   -\frac{e^{i \phi } \gamma  (\gamma +1)}{\gamma  (8 \gamma +27)+27} & 0
   \\
 0 & -\frac{e^{-i \phi } \gamma  (\gamma +1)}{\gamma  (8 \gamma +27)+27} &
   \frac{(\gamma +2) (\gamma +3)}{\gamma  (8 \gamma +27)+27} & 0 \\
 0 & 0 & 0 & \frac{2 (\gamma  (3 \gamma +8)+6)}{\gamma  (8 \gamma +27)+27}
   \\
\end{array}
\right)
\end{equation}


\begin{figure}[ht]
  \centering
    \subfloat[Probability density of the phase difference $\Delta \theta$ for given coupling phase $\phi$.]
        {{\includegraphics[width=0.33\textwidth]{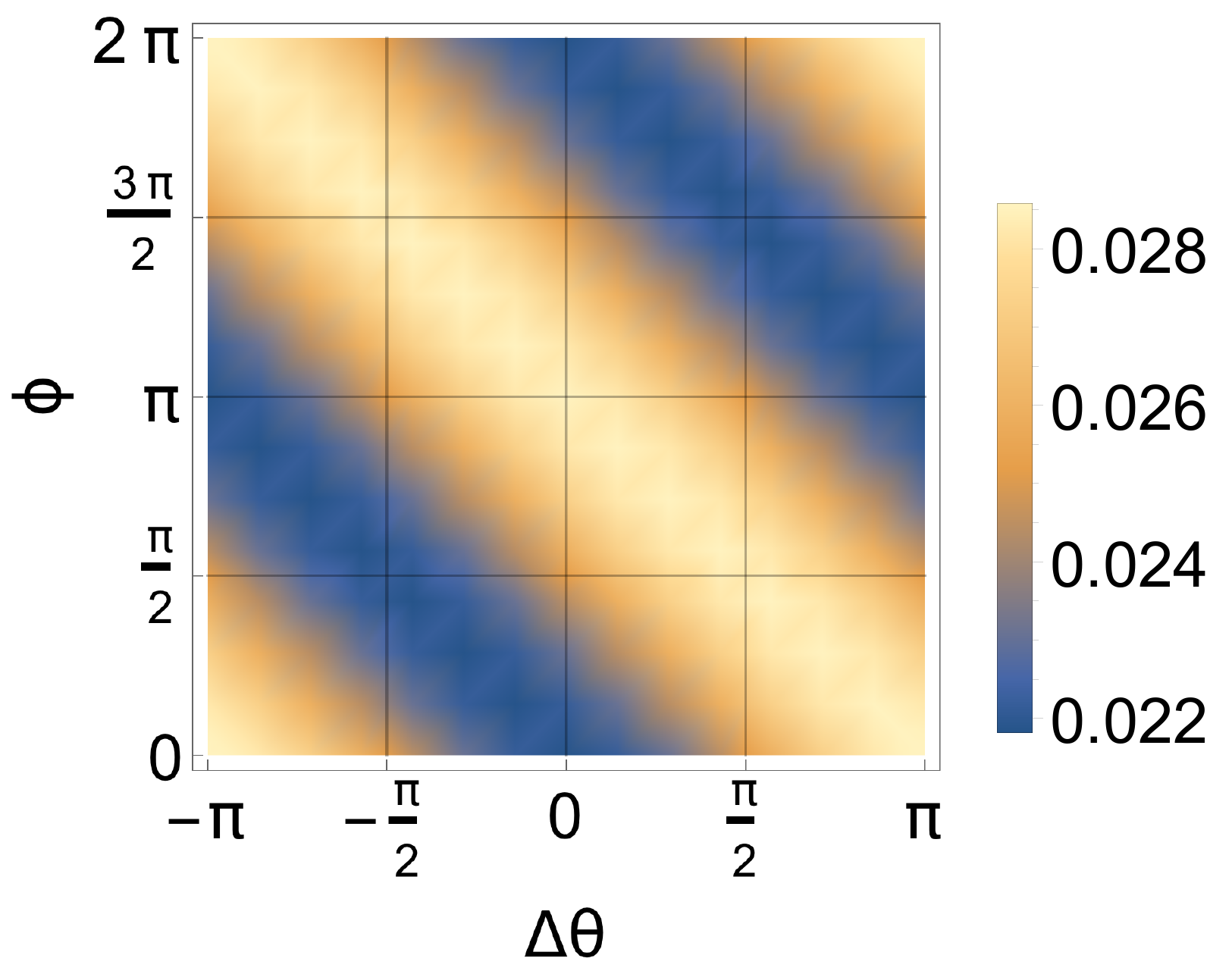}}
  \label{fig:CouplingPhaseSynchronizationPhase}
    }
    \qquad
  \centering
    \subfloat[Expectation value of the momentum correlation function given in Eq.~(\ref{eq:pxp_expct}) for $\kappa=1$. Initially there is an approximately linear increase.]
        {{\includegraphics[width=0.57\textwidth]{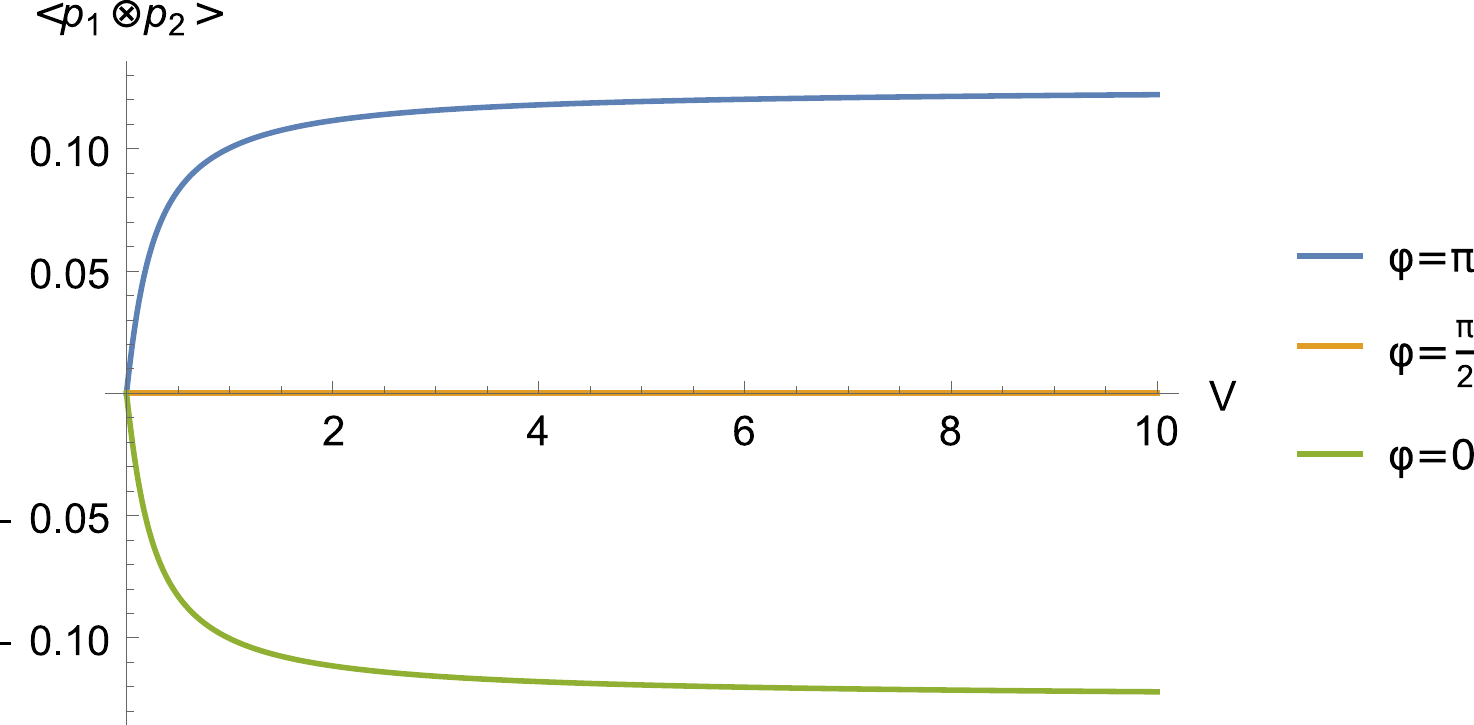}}
  \label{fig:Pxp_over_V}}
    \caption{Different measures of synchronization for two coupled oscillators.}
\end{figure}

\subsection{Influence of the coupling phase}

The Wigner function can be defined in the same way for two oscillators. However, instead of using $x_1,x_2,p_1,$ and $p_2$ as arguments
we make a transformation to polar coordinates with 
\begin{equation}
r_je^{i\theta_j} = \sqrt{2}(x_j + ip_j)  \qquad\qquad (j=1,2)
\end{equation}
as described in the Appendix of Ref.~\cite{Lee2014}.
Using this polar representation 
we can ignore the magnitude and only look at
the probability distribution for the phase difference $\Delta\theta=\theta_2-\theta_1$ of the two oscillators.

We find that the probability is maximal for
 $\Delta\theta=\phi-\pi$.
This can be seen in Fig.~\ref{fig:CouplingPhaseSynchronizationPhase}
where the probability density of the synchronization phase $\Delta\theta$
is plotted in dependence of the coupling phase $\phi$
for fixed values  $V=10$ and $\kappa=0.1$.

Synchronization can also be detected by measuring the momentum correlation function which is defined as the expectation value 
\begin{equation}
   \left<p_1\otimes p_2\right>=
   \Tr[(p_1\otimes p_2) \rho]=
  -\frac{\gamma  (\gamma +1) \cos (\phi )}{\gamma  (8 \gamma +27)+27}\,.
    \label{eq:pxp_expct}
    \label{correlator}
\end{equation}
This correlator depends on the coupling phase $\phi$ and vanishes if the oscillators are coupled out of phase (meaning that $\phi=\frac{\pi}{2}$). Moreover, as can be seen in Fig.~\ref{fig:Pxp_over_V} and Eq. (\ref{correlator}) it is bound by  $|\left<p_1\otimes p_2\right>| \leq \frac18$. Obviously, the behavior of this correlator is in agreement with the previous findings obtained by the Wigner function.

\subsection{Entanglement}
\begin{figure}[ht]
  \centering
  \includegraphics[width=0.7\textwidth]{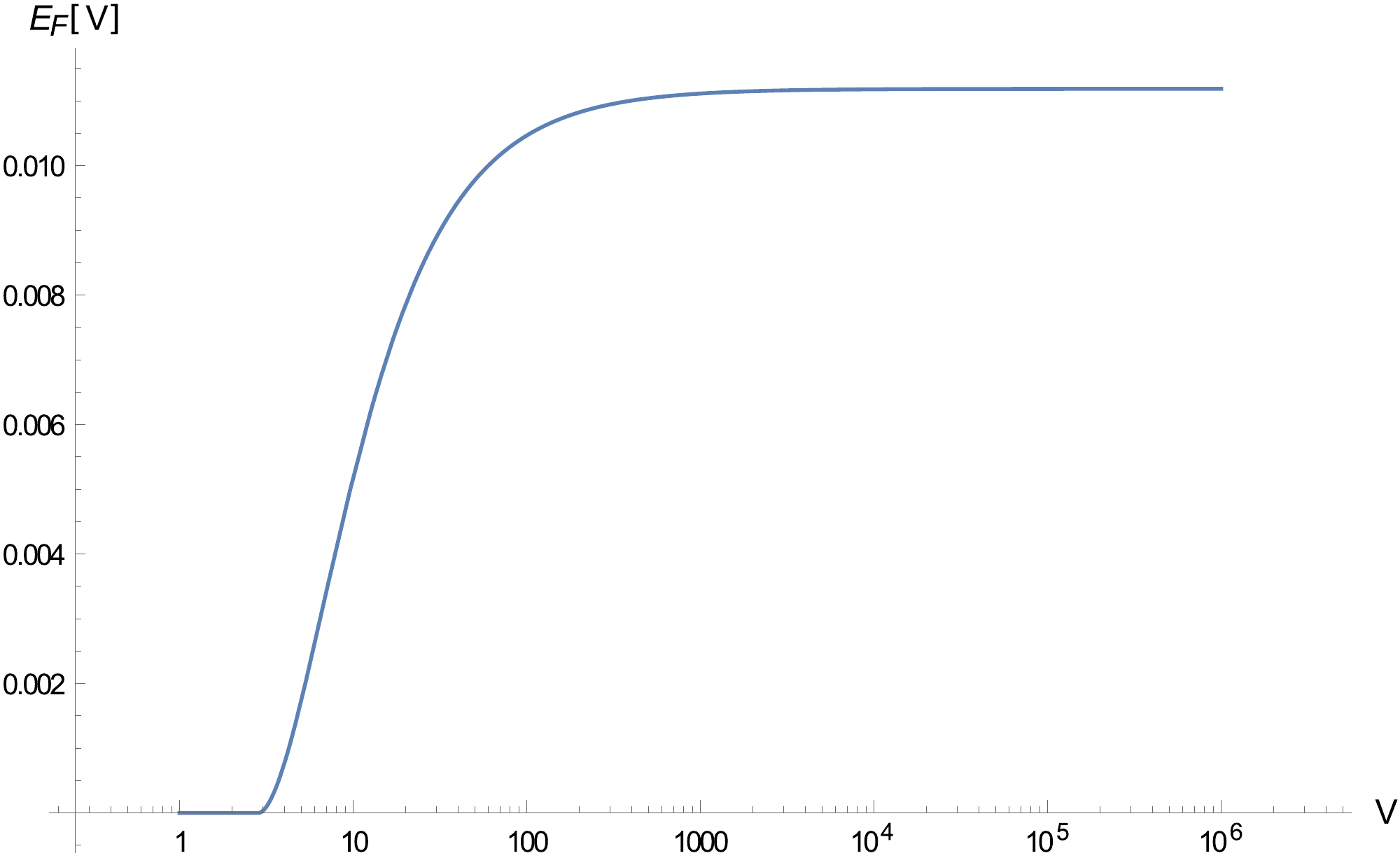}
  \caption[]{Entanglement of formation for two coupled oscillators (steady state solution of
      equation~(\ref{eq:TwoCoupledVdP})) plotted over coupling strength $V$.}
  \label{fig:Entanglement_of_formation}
\end{figure}

An important property of coupled quantum mechanical systems is the emergence of entanglement.
For two two-level systems (i.e. qubits) the common entanglement measure is the entanglement of formation~\cite{Wootters98}.
It is defined by the formulas
\begin{eqnarray}
    &E_F(\rho)=S\left[\frac{1}{2}(1+\sqrt{1-C^2(\rho)})\right]\nonumber\\[0.2em]
    \textnormal{with}\qquad
    &S(x)=-x\log_2x-(1-x)\log_2(1-x) \\[0.2em]
    \textnormal{and}\qquad
    &C(\rho)=\max(0,\lambda_1-\lambda_2-\lambda_3-\lambda_4)\,, \nonumber\\ \nonumber
\end{eqnarray}
where $\{\lambda_1,\lambda_2,\lambda_3,\lambda_4\}$ are the eigenvalues of the operator $\rho(\sigma^y\otimes\sigma^y)\rho^*(\sigma^y\otimes\sigma^y)$ sorted by magnitude in descending order. In the present case we have
\begin{eqnarray}
\lambda_1 &=& \frac{4 (\gamma  (\gamma +3)+3)^2}{(\gamma  (8 \gamma +27)+27)^2} \\
\lambda_2 = \lambda_3 &=&\frac{2 (\gamma +3) (\gamma  (3 \gamma +8)+6)}{(\gamma  (8 \gamma
   +27)+27)^2}\\
\lambda_4 &=& \frac{4 (2 \gamma +3)^2}{(\gamma  (8 \gamma +27)+27)^2}
\end{eqnarray}
so that $\lambda_1-\lambda_2-\lambda_3-\lambda_4=\frac{4 (\gamma +3) \left(\gamma ^3-6 \gamma -6\right)}{(\gamma  (8 \gamma+27)+27)^2}$. Inserting this expression into the formulas listed above, the entanglement of formation can be computed directly. The result is shown Figure~\ref{fig:Entanglement_of_formation}. As can be seen, the entanglement vanishes for small values of $\gamma$. The onset of entanglement happens at a coupling ratio of $\gamma=2^{1/3}+2^{2/3}\approx2.847$. For large values of $\gamma$ the entanglement reaches an upper limit of $E_F(\rho)=S[\frac12+\frac{\sqrt{255}}{32}]\approx0.01118$. This means that even for very large coupling strength the entanglement is weak. A maximally entangled two-qubit state, e.g. a Bell state, has $E_F=1$.

\subsection{Eigenvalue spectrum of the reduced density matrices}

\begin{figure}[h]
  \centering
  \includegraphics[width=0.7\textwidth]{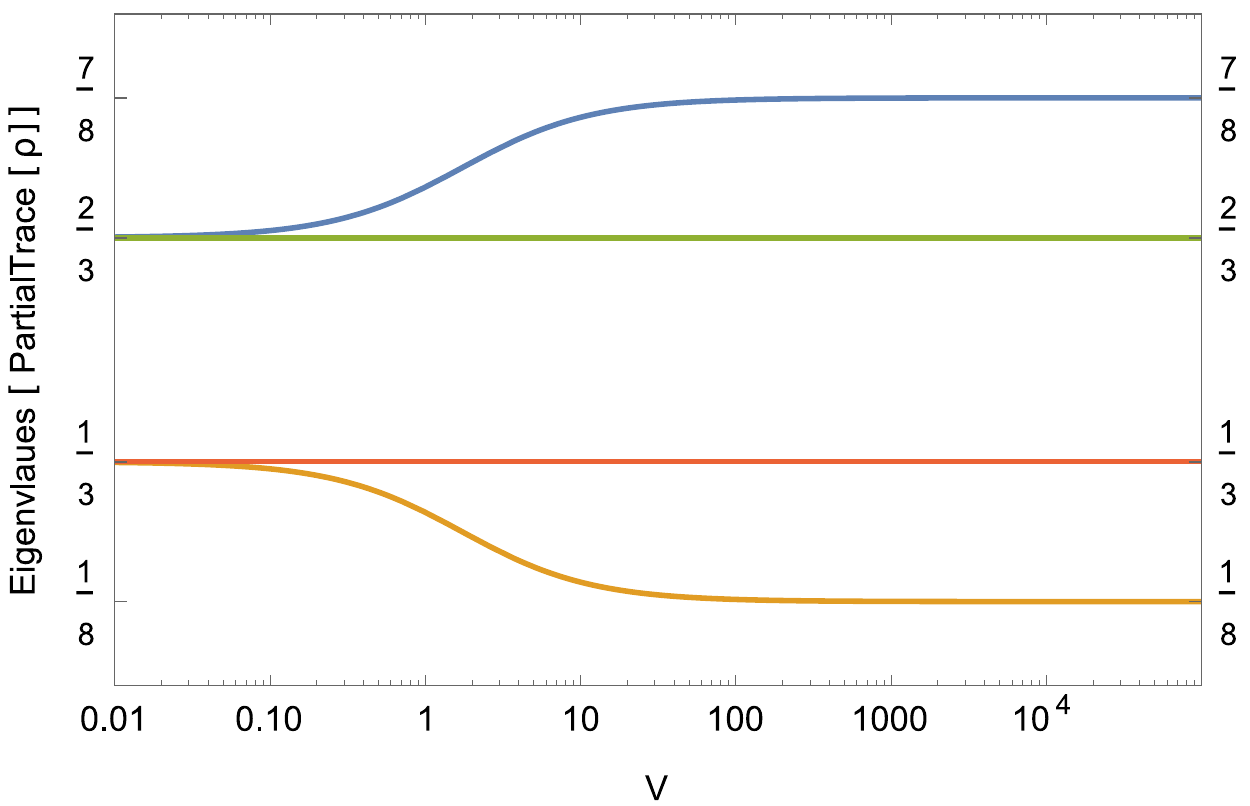}
    \caption[]{Spectrum of the partial density matrix (equation~(\ref{eq:Partial_rho})). The
    horizontal lines are the eigenvalues of the uncoupled oscillator.}
  \label{fig:Eigenvalue_Transition}
\end{figure}

In order to study the influence of mutual coupling on a single oscillator, we compute the reduced density matrix:
\begin{equation}
\left(\begin{array}{cc}
    \frac{7V^2 +21V\kappa +18\kappa^2}{8V^2 +27V\kappa +27\kappa^2} &
    0\\
    0&
    \frac{(V+3\kappa)^2}{8V^2 +27V\kappa +27\kappa^2}\\
\end{array}\right).\label{eq:Partial_rho}
\end{equation}
In the limit $V \to 0$ this reproduces our single oscillator result
(compare section~\ref{sec:SingleVdP}).
For large values of $V$ the eigenvalues of the single oscillator
tend to $\frac{7}{8}$ and $\frac{1}{8}$, as shown in Fig.~\ref{fig:Eigenvalue_Transition}.
The parameter $\kappa$ defines how quickly this transition happens.

\section{Conclusions}

In this work we have studied the synchronization properties of two self-sustained quantum oscillators coupled by an interaction term involving a phase shift. We confirm that the interaction leads to phase correlations between the systems. In particular, these correlations depend on the phase shift in the interaction term. These correlations have been visualized by using Wigner functions and evaluated explicitly. The correlations are a property of the stationary state, meaning that they are not destroyed by decoherence. Thus the system shows a stable correlated behavior although it does not synchronize perfectly in the classical sense. We have also analyzed the entanglement of the oscillators as a function of the coupling strength in the stationary state. While correlations are present for any value of $V>0$ we find that entanglement only sets in at a well-define threshold of the coupling strength.

It would be interesting to extend this analysis to more than two oscillators. For example, in the classical case it is known that four oscillators synchronize partially in subgroups, and it would be interesting to find out whether this feature survives in the quantum limit. Moreover, it would be interesting to understand the meaning of the eigenvalues of the reduced density matrix.

\appendix

\section{Alternative derivation from a spin system in a spin bath}\label{sec:SpinSpin}

Eq.~\eqr{eq:MasterOneVdP} can also be obtained in a different way by following Englert and Morigi~\cite{Englert2002}, who derived a master equation very similar to the
quantum van der Pol master equation, however, lacking the
square in the second Lindblad term.

If we perform an analogous  derivation with
an oscillator Hamiltonian of a spin system in a spin bath
it gives us right away the master equation in the quantum limit~\eqr{eq:MasterOneVdP}.
Starting point of the derivation a single spin representing the system $s$ and
an arbitrarily large number of spins representing the environment $e$. 

For the interaction part we make the following assumption.
We can imagine that at different times individual bath atoms touch the system and interact with it.
Then the system evolves freely again until the next bath spin arrives.
We assume that only one of the environment spins
interacts with the system spin at any time. Therefore, we consider the Hamiltonian
\begin{equation} \label{eq:Hamiltonian}
    H=  \underbrace{\omega \left( \sigSD \sigS + \sigED \sigE \right)}_{H_\text{free}}
        + \underbrace{g    \left( \sigED \sigS + \sigSD \sigE \right)}_{H_\text{int}},
\end{equation}
where only one of the environment spins appears in the interaction part.

In the end, when we put together the master equation, we will
account for the different bath spins that interact with the bath at different times
by performing an appropriate average.

In the literature on quantum optics a similar Hamiltonian is studied under the name
``resonant Jaynes-Cummings interaction in the rotating-wave approximation'',
where the system is modeled by a harmonic oscillator instead of a spin.
The first term $H_\text{free}$ describes the free evolution of the two spins.
For simplicity we assume that they both have the same frequency $\omega$.
The coupling between the two spins is described by the second (exchange) term $H_\text{int}$
controlled by the coupling strength $g$.
The time evolution of the density matrix $\rho$ is given by
von Neumann's equation
\begin{equation}\label{eq:Neumann}
   \frac{\partial}{\partial t} \rho_t = i \left[ \rho_t , H \right].
\end{equation}
Let us now look at a timespan $[t,t+\tau]$ during which the interaction between the system
and a single bath electron takes place.
For this we first note that the two terms in the Hamiltonian~\eqr{eq:Hamiltonian}
$H_\text{free}$ and $H_\text{int}$ commute:
\begin{eqnarray}\label{eq:commutation}
\fl
    \left[ H_\text{free} , H_\text{int} \right] &=&
    \left[
    \omega \left( \sigSD \sigS + \sigED \sigE \right),
     g \left( \sigED \sigS + \sigSD \sigE \right)
    \right]\\ \nonumber &=& \omega g \left(
    \left[ \sigSD \sigS, \sigED \sigS \right] +
    \left[ \sigSD \sigS, \sigSD \sigE \right] +
    \left[ \sigED \sigE, \sigED \sigS \right] +
    \left[ \sigED \sigE, \sigSD \sigE  \right]
    \right)\\ \nonumber &=& \omega g \left(
    \sigED \left[ \sigSD , \sigS \right] \sigS +
    \sigSD \left[ \sigS , \sigSD \right] \sigE +
    \sigED \left[ \sigE , \sigED \right] \sigS +
    \sigSD \left[ \sigED , \sigE  \right] \sigE
    \right)\\ \nonumber &=& \omega g \left(
    -\sigED \sigS +
    \sigSD \sigE +
    \sigED \sigS -
    \sigSD \sigE
    \right) = 0 \,. \nonumber
\end{eqnarray}
Under this condition the state of the total system at time $t+\tau$ reads
\begin{eqnarray}\label{eq:SplitTimEv}
    \rho_{t+\tau} &=& e^{ -i H \tau} \rho_t e^{ i H \tau} \\
    &=& e^{ -i H_\text{free} \tau} \left[ e^{ -i H_\text{int} \tau}
    \rho_t e^{ i H_\text{int} \tau} \right] e^{ i H_\text{free} \tau} \nonumber \\
    &=&  e^{ -i H_\text{free} \tau} \left[ e^{-i \gamma \phi}
    \rho_t e^{i \gamma \phi} \right] e^{ i H_\text{free} \tau},\nonumber
\end{eqnarray}
where we introduced the notations $\gamma = \left( \sigED \sigS + \sigSD \sigE \right)$ and  $\phi = g \tau$.
We now rewrite the term
$\left[ e^{-i \gamma \phi} \rho_t e^{i \gamma \phi} \right]$
which amounts for the change introduced by the coupling.
The first exponential can be written as
\begin{eqnarray}\label{eq:GammaSin}
    e^{i \phi \gamma} &= \cos\!\left(\phi \gamma\right) + i \sin\!\left(\phi \gamma\right)
    = \cos\!\left(\phi \sqrt{\gamma^2}\right)
    + i \gamma \frac{\sin\!\left(\phi \sqrt{\gamma^2}\right)}{\sqrt{\gamma^2}}.
\end{eqnarray}
It is convenient to use this form because we can apply some general rules, namely
\begin{eqnarray}
\label{firstblock1}
    \left\{ \sigma^\dagger , \sigma \right\} &=& \mathbf{1} \\
    \label{firstblock2}
    \sigma^2 = \left( \sigma^\dagger \right)^2 &=& 0 \\
    \label{secondblock}
    \sigma f\! \left( \sigma^\dagger \sigma \right) &=& \sigma f\! \left( \mathbf{1} \right) \\
    \sigma^\dagger f\! \left( \sigma^\dagger \sigma \right) &=& \sigma^\dagger f\! \left( 0 \right)\\
    \sigma^\dagger f\! \left( \sigma \sigma^\dagger  \right) &=& \sigma^\dagger f\! \left(
\mathbf{1} \right) \\
    \sigma f\! \left( \sigma \sigma^\dagger  \right) &=& \sigma f\! \left( 0 \right)
\end{eqnarray}
These rules apply to functions $f$ of operators in general.
Applying the basic relations of fermionic operators (\ref{firstblock1}) and (\ref{firstblock2})  we can calculate
$ \gamma^2 = \sigED \sigE \sigS \sigSD + \sigE \sigED
    \sigSD \sigS $.
Let us use the rest of the rules to transform a function $ F\! \left( \gamma^2 \right) $
into a suitable form, which we can then employ for the next step of the derivation.
\begin{eqnarray}
    F\! \left( \gamma^2 \right) &=
    \underbrace{ \left( \sigED \sigE + \sigE \sigED \right) }_{ = \mathbf{1} }
    F\! \left(\sigED \sigE \sigS \sigSD + \sigE \sigED
    \sigSD \sigS \right) \\\nonumber &=
    \sigED \sigE~F\! \left( \mathbf{1} \otimes \sigS \sigSD + 0 \cdot \sigSD \sigS \right) +
    \sigE \sigED~F\! \left( 0 \cdot \sigS \sigSD + \mathbf{1} \otimes \sigSD \sigS \right) \\\nonumber &=
    \sigED \sigE~F\! \left( \sigS \sigSD \right) +
    \sigE \sigED~F\! \left( \sigSD \sigS \right)\\
    \gamma F\! \left( \gamma^2 \right) &=
    \left( \sigED \sigS + \sigSD \sigE \right)
    F\! \left( \sigED \sigE \sigS \sigSD + \sigE \sigED
    \sigSD \sigS \right) \\ &=
    \sigED~F\! \left( \sigS \sigSD \right) \sigS +
    \sigE\sigSD~F\! \left( \sigS \sigSD \right) \nonumber
\end{eqnarray}
We can apply this to~\eqr{eq:GammaSin}:
\begin{eqnarray}\label{eq:GammaSinExp}
    &\cos\!\left(\phi \sqrt{\gamma^2}\right)
    + i \gamma \frac{\sin\!\left(\phi \sqrt{\gamma^2}\right)}{\sqrt{\gamma^2}} = \\ \nonumber
    &\sigED \sigE~\underbrace{\cos\!\left(\phi \sqrt{\sigS \sigSD}\right)}_{C}
    + \sigE \sigED~\underbrace{\cos\!\left(\phi \sqrt{\sigSD \sigS}\right)}_{\tilde{C}} \\
    &+ i \sigE~\underbrace{ \sigSD \frac{\sin\!\left(\phi \sqrt{\sigS \sigSD}\right)}{\sqrt{
    \sigS \sigSD}}}_{S}
    + i \sigED \underbrace{ \frac{\sin\!\left(\phi \sqrt{\sigS \sigSD}\right)}{\sqrt{\sigS \sigSD
    }}\sigS}_{S^\dagger} \nonumber
\end{eqnarray}
In the following we use the abbreviations ($C$, $\tilde{C}$, $S^\dagger$ and $S$)
introduced in this equation~\eqr{eq:GammaSinExp}.

Up to here our derivation depends only on algebraic relations
and is independent of the choice of the density matrix.
To be able to proceed with the derivation,
we now choose a representation for the $\sigE$-operators.

Consider a vector representation of the bath atom where
    $\ket{0} = (0,1)^T$ and
    $\ket{1} = (1,0)^T$.
Using this representation the $\sigE$-operators are actually $2\times2$-matrices.

With this notation~\eqr{eq:GammaSinExp} simplifies to
\begin{eqnarray}\label{eq:GammaSinMat}
    e^{i \phi \gamma} =
    \cos\!\left(\phi \sqrt{\gamma^2} \right)
    + i \gamma \frac{\sin\! \left( \phi \sqrt{\gamma^2} \right)}{\sqrt{\gamma^2}} =
    \left(\begin{array}{c c} C & i S^\dagger  \\ i S & \tilde{C} \end{array}\right),
\end{eqnarray}
where the final matrix is actually a $4\times4$ matrix.

Next we make some assumptions about the density matrix.
Before the interaction (at time $t$), the system and bath particle are independent.
In mathematical terms this means that they are separable
and the total density matrix can be written as a tensor product
of the two contributing terms $\rho_t = \rho_t^{(s)}\otimes\rho_t^{(e)}$.
Ultimately we are only interested in the reduced density matrix of the system.
Tracing out the bath particle we obtain
\begin{eqnarray}\label{eq:SplitTimEvTr}
    \rho_{ t + \tau }^{(s)} =&
    \Tr_e \left[ e^{-i H \tau} \rho_t e^{i H \tau} \right] \\
    =&  e^{-i \omega \sigSD \sigS \tau} \Tr_e \left[ e^{-i \gamma \phi}
    \rho_t^{(s)} \otimes \rho_t^{(e)} e^{i \gamma \phi} \right] e^{i \omega \sigSD \sigS \tau}\nonumber\\
    =&  e^{-i \omega \sigSD \sigS \tau}
    \Tr_e \left[ \left(\begin{array}{c c} C & -i S^\dagger  \\ -i S & \tilde{C} \end{array}\right)
    \rho_t^{(s)} \otimes \rho_t^{(e)} \left(\begin{array}{c c} C & i S^\dagger  \\ i S & \tilde{C} \end{array}\right)
    \right] e^{i \omega \sigSD \sigS \tau}.\nonumber
\end{eqnarray}
At this point we recall our assumption about the bath spin that we made in the beginning,
that it is one of many spin particles in the bath.
Now we specify the state of this bath spin at the time when the interaction starts.
It can either be in the ground state
    $\rho_t^{(e)}=\left( \begin{array}{c c} 0 & 0 \\ 0 & 1 \end{array}\right)$
    or the excited state
$\rho_t^{(e)}=\left( \begin{array}{c c} 1 & 0 \\ 0 & 0 \end{array}\right)$.
The time evolution~\eqr{eq:SplitTimEvTr} yields a different result for each of these states.
Therefore we do the next steps for both cases in parallel.

For a particle arriving in the ground state we have
\begin{eqnarray*}
    \left(\begin{array}{c c} C & -i S^\dagger  \\ -i S & \tilde{C} \end{array}\right)
    \left(\begin{array}{c c} 0 & 0 \\ 0 & \rho_t^{(s)} \end{array}\right)
    \left(\begin{array}{c c} C & i S^\dagger  \\ i S & \tilde{C} \end{array}\right)=
    \left(\begin{array}{c c} S^\dagger \rho_t^{(s)} S & -i S^\dagger \rho_t^{(s)} \tilde{C} \\
    i \tilde{C} \rho_t^{(s)} S & \tilde{C} \rho_t^{(s)} \tilde{C}\end{array}\right)
\end{eqnarray*}
and for the excited state we have
\begin{eqnarray*}
    \left(\begin{array}{c c} C & -i S^\dagger  \\ -i S & \tilde{C} \end{array}\right)
    \left(\begin{array}{c c} \rho_t^{(s)} & 0 \\ 0 & 0 \end{array}\right)
    \left(\begin{array}{c c} C & i S^\dagger  \\ i S & \tilde{C} \end{array}\right)=
    \left(\begin{array}{c c} C \rho_t^{(s)} C & i C \rho_t^{(s)} S^\dagger \\
    -i S \rho_t^{(s)} C & S \rho_t^{(s)} S^\dagger  \end{array}\right).
\end{eqnarray*}
We can finally calculate the trace by inserting this into the time
evolution~\eqr{eq:SplitTimEvTr}
for the ground state
\begin{eqnarray}\label{eq:SplitTimEvTrGs}
    \rho_{ t + \tau }^{(s)}
    &= e^{-i \omega \sigSD \sigS \tau}
    \Tr_e \left[
    \left(\begin{array}{c c} S^\dagger \rho_t^{(s)} S & -i S^\dagger \rho_t^{(s)} \tilde{C} \\
    i \tilde{C} \rho_t^{(s)} S & \tilde{C} \rho_t^{(s)} \tilde{C}\end{array}\right)
    \right]
    e^{i \omega \sigSD \sigS \tau}\\ \nonumber
    &=  e^{-i \omega \sigSD \sigS \tau}
    \left[ S^\dagger \rho_t^{(s)} S +  \tilde{C} \rho_t^{(s)} \tilde{C} \right]
    e^{i \omega \sigSD \sigS \tau}\\ \nonumber
\end{eqnarray}
and the excited state
\begin{eqnarray}\label{eq:SplitTimEvTrEs}
    \rho_{ t + \tau }^{(s)}
    &= e^{-i \omega \sigSD \sigS \tau}
    \Tr_e \left[
    \left(\begin{array}{c c} C \rho_t^{(s)} C & i C \rho_t^{(s)} S^\dagger \\
    -i S \rho_t^{(s)} C & S \rho_t^{(s)} S^\dagger  \end{array}\right)
    \right]
    e^{i \omega \sigSD \sigS \tau}\\ \nonumber
    &=  e^{-i \omega \sigSD \sigS \tau}
    \left[  C \rho_t^{(s)} C + S \rho_t^{(s)} S^\dagger \right]
    e^{i \omega \sigSD \sigS \tau}.\\ \nonumber
\end{eqnarray}
To proceed further we expand these terms around $\phi \approx 0$ up to
$2$nd order in $\phi$.
This makes our derivation only valid for weak interactions.
We pick up the derivation in~\eqr{eq:SplitTimEvTrGs} and explicitly write out the
terms $C$, $\tilde{C}$, $S$ and $S^\dagger$, before we expand the sine and cosine terms in $\phi$.

\begin{eqnarray}\label{eq:TimEvGs}
    \rho_{ t + \tau }^{(s)}
    =  e^{-i \omega \sigSD \sigS \tau}
    &\left[ S^\dagger \rho_t^{(s)} S +  \tilde{C} \rho_t^{(s)} \tilde{C} \right]
    e^{i \omega \sigSD \sigS \tau}\\ \nonumber
    =  e^{-i \omega \sigSD \sigS \tau}
    &\left[  \frac{\sin\!\left(\phi \sqrt{\sigS \sigSD}\right)}{\sqrt{\sigS \sigSD }}\sigS
    \rho_t^{(s)}\right.
    \sigSD \frac{\sin\!\left(\phi \sqrt{\sigS \sigSD}\right)}{\sqrt{ \sigS \sigSD}} \\ \nonumber
    &+ \left.\vphantom{\frac{\sin\!\left(\phi \sqrt{\sigS \sigSD}\right)}{\sqrt{\sigS \sigSD }}}
    \cos\!\left(\phi \sqrt{\sigSD \sigS}\right)
    \rho_t^{(s)} \cos\!\left(\phi \sqrt{\sigSD \sigS}\right)\right]
    e^{i \omega \sigSD \sigS \tau}\\ \nonumber
    =  e^{-i \omega \sigSD \sigS \tau}
    &\left[ \vphantom{\frac{1}{2}} \sigS \phi \rho_t^{(s)} \right. \phi \sigSD \\ \nonumber
    & + \left( 1 - \frac{1}{2} \phi^2 \sigSD \sigS \right)\left.
    \rho_t^{(s)} \left( 1 - \frac{1}{2} \phi^2 \sigSD \sigS \right) \right]
    e^{i \omega \sigSD \sigS \tau}\\ \nonumber
    =  e^{-i \omega \sigSD \sigS \tau}
    &\left[ \frac{1}{2} \phi^2 \left( 2 \sigS \rho_t^{(s)} \sigSD
    -\sigSD \sigS \rho_t^{(s)}
    - \rho_t^{(s)} \sigSD \sigS \right) + \rho_t^{(s)} \right]
    e^{i \omega \sigSD \sigS \tau}.\\ \nonumber
\end{eqnarray}
Applying the same process for the interaction with bath particles initially in the excited state
we get a time evolution
\begin{eqnarray}\label{eq:TimEvEs}
    \rho_{ t + \tau }^{(s)}
    =  e^{-i \omega \sigSD \sigS \tau}
    &\left[  C \rho_t^{(s)} C + S \rho_t^{(s)} S^\dagger \right]
    e^{i \omega \sigSD \sigS \tau}\\ \nonumber
    =  e^{-i \omega \sigSD \sigS \tau}
    &\left[ \frac{1}{2} \phi^2 \left( 2 \sigSD \rho_t^{(s)} \sigS
    -\sigS \sigSD \rho_t^{(s)}
    - \rho_t^{(s)} \sigS \sigSD \right) + \rho_t^{(s)} \right]
    e^{i \omega \sigSD \sigS \tau}.\\ \nonumber
\end{eqnarray}
Note that the $0$th order term in both equations is the free evolution of $\rho_t^{(s)}$ without the
bath.
The change induced by the bath alone is
\begin{eqnarray}\label{eq:TimEvGsDelta}
    \Delta_1 \rho_t^{(s)}
    = \frac{1}{2} \phi^2 \left( 2 \sigS \rho_t^{(s)} \sigSD
    -\sigSD \sigS \rho_t^{(s)}
    - \rho_t^{(s)} \sigSD \sigS \right)
\end{eqnarray}
or
\begin{eqnarray}\label{eq:TimEvEsDelta}
    \Delta_2 \rho_t^{(s)}
    = \frac{1}{2} \phi^2 \left( 2 \sigSD \rho_t^{(s)} \sigS
    -\sigS \sigSD \rho_t^{(s)}
    - \rho_t^{(s)} \sigS \sigSD \right).
\end{eqnarray}
We can further assume that the arrival times of these bath spins are independent,
this means they follow Poissonian statistics.
The average time between the arrival of two bath spins (i.e. the inverse of the arrival rate)
is assumed to be much larger than the interaction time $\tau$.
This is necessary to ensure that not two bath spins interact with the system at the same time.
If we specify an arrival rate $\kappa$ for ground state atoms
and $\lambda$ for excited state atoms we can now write~\eqr{eq:Neumann} for $\rho_t^{(s)}$ :
\begin{eqnarray}\label{eq:SpinSpinResult}
    \frac{\partial}{\partial t} \rho_t^{(s)} & = i \left[ \rho_t^{(s)} , H \right]
        & \\\nonumber
    & = \left[ \rho_t^{(s)} , \omega \sigSD \sigS \right]
        & + \frac{1}{2} \phi^2 \kappa \left[ 2 \sigS \rho_t^{(s)} \sigSD
        - \sigSD \sigS \rho_t^{(s)}
        - \rho_t^{(s)} \sigSD \sigS \right] \\\nonumber
    &   & + \frac{1}{2} \phi^2 \lambda \left[ 2 \sigSD \rho_t^{(s)} \sigS
        - \sigS \sigSD \rho_t^{(s)}
        - \rho_t^{(s)} \sigS \sigSD \right]
\end{eqnarray}
This is the same as Eq.~\ref{eq:MasterOneVdP} with 
just a little difference in notation:
Here $\sigma^\dagger$ and $\sigma$ are used instead of $\sigma^+$ and $\sigma^-$.

\section*{References}

\label{Bibliography}

\bibliographystyle{unsrt}

\bibliography{Bibliography}

\begin{thebibliography}{10}

\bibitem{Blekhman88}
Blekhman~L. I.
\newblock Synchronization in science and technology.
\newblock 1988.

\bibitem{Boccaletti02}
S.~Boccaletti, J.~Kurths, G.~Osipov, D.L. Valladares, and C.S. Zhou.
\newblock The synchronization of chaotic systems.
\newblock {\em Physics Reports}, 366(1–2):1 -- 101, 2002.

\bibitem{Schlosshauer2008}
Maximilian~A. Schlosshauer.
\newblock {\em Decoherence and the {Quantum}-to-{Classical} {Transition}}.
\newblock Springer, Berlin ; London, April 2008.

\bibitem{Giorgi2014}
G.~Giorgi, F.~Plastina, G.~Francica, and R.~Zambrini.
\newblock Spontaneous synchronization and quantum correlation dynamics of open
  spin systems.
\newblock {\em Physical Review A}, 88(4):042115, October 2013.

\bibitem{VanDerPol34}
Balth and van~der Pol.
\newblock The nonlinear theory of electric oscillations.
\newblock {\em Proc. Inst. Radio Engineers}, 22:1051--1086, 1934.

\bibitem{Nana06}
B.~Nana and P.~Woafo.
\newblock Synchronization in a ring of four mutually coupled van der pol
  oscillators: Theory and experiment.
\newblock {\em Phys. Rev. E}, 74:046213, Oct 2006.

\bibitem{Lee2013}
Tony~E. Lee and H.~R. Sadeghpour.
\newblock Quantum {Synchronization} of {Quantum} van der {Pol} {Oscillators}
  with {Trapped} {Ions}.
\newblock {\em Physical Review Letters}, 111(23):234101, December 2013.

\bibitem{Walter2014}
Stefan Walter, Andreas Nunnenkamp, and Christoph Bruder.
\newblock Quantum synchronization of two {Van} der {Pol} oscillators.
\newblock {\em Annalen der Physik}, pages n/a--n/a, August 2014.

\bibitem{Lee2014}
Tony~E. Lee, Ching-Kit Chan, and Shenshen Wang.
\newblock Entanglement tongue and quantum synchronization of disordered
  oscillators.
\newblock {\em Physical Review E}, 89(2):022913, February 2014.

\bibitem{Englert2002}
Berthold-Georg Englert and Giovanna Morigi.
\newblock Five {Lectures} on {Dissipative} {Master} {Equations}.
\newblock In Andreas Buchleitner and Klaus Hornberger, editors, {\em Coherent
  {Evolution} in {Noisy} {Environments}}, number 611 in Lecture {Notes} in
  {Physics}, pages 55--106. Springer Berlin Heidelberg, January 2002.

\bibitem{Leonhardt2005}
Ulf Leonhardt.
\newblock {\em Measuring the {Quantum} {State} of {Light}}.
\newblock Cambridge University Press, Cambridge, N.Y, November 2005.

\bibitem{Wootters98}
William~K. Wootters.
\newblock Entanglement of formation of an arbitrary state of two qubits.
\newblock 80(10):2245--2248.

\end{thebibliography}
\end{document}